# PARALLEL IMPLEMENTATIONS OF THE JACOBI LINEAR ALGEBRAIC SYSTEMS SOLVER

ATHANASIOS MARGARIS[1], STAVROS SOURAVLAS[1], MANOS ROUMELIOTIS[1]

[1]Department of Applied Informatics, University of Macedonia
Egnatia 156 Street, GR 540 06, Thessaloniki, Greece
email: amarg@uom.gr; sourstav@uom.gr; manos@uom.gr

The objective of this research is to construct parallel implementations of the Jacobi algorithm used for the solution of linear algebraic systems, to measure their speedup with respect to the serial case and to compare each other, regarding their efficiency. The programming paradigm used in this implementation is the message passing model, while, the MPI implementation used is the MPICH implementation of the Argonne National Laboratory.

## 1  INTRODUCTION

The Message Passing Interface (MPI) offers a portable efficient and inexpensive method of implementing parallel algorithms in clusters of workstations. It can be used for a variety of applications, but it is most effective for the parallelization of problems that can be split into many processes. The Jacobi iterative algorithms are very good candidates for parallelization on an MPI cluster due to being very computationally intensive and inherently divisible into parallel tasks.

Several efforts have been focused on parallelizing Jacobi algorithms. Luke and Park (Luce & Park, 1989) consider two parallel Jacobi algorithms for computing the singular value decomposition of an nxn matrix. By relating the algorithms to the cyclic-by-rows Jacobi method, they prove convergence of one algorithm for odd n and of another for any n. Zhou and Brent (Zhou & Brent, 1995) show the importance of the sorting of column norms in each sweep for one-sided Jacobi SVD computation. They describe two parallel Jacobi orderings. These orderings generate n(n-1)/2 different index pairs and sort column norms at the same time. The one-sided Jacobi SVD algorithm using these parallel orderings converges in about the same number of sweeps as the sequential cyclic Jacobi algorithm. Gimenez et. al (Gimenez et. al, 1995, Royo et. al, 1998) study the parallelization of the Jacobi method to solve the symmetric eigenvalue problem on a mesh of processors and hypercubes. They show how matrix symmetry can be exploited on a logical mesh of processors to obtain high scalability. Londre and Rhee (Londre & Rhee, 2005) study the numerical stability of the parallel Jacobi method for computing the singular values and singular subspaces of an invertible upper triangular matrix. Coope and Macklem (Coope & Macklem, 2005) present how to efficiently use parallel and distributed computing platforms in solving derivative-free optimization problems with the Jacobi algorithm. Convergence is achieved by introducing an elementary trust region subproblem at synchronization steps in the algorithm. This has the added advantage of handling negative curvature very conveniently. Acacio et. al (Acacio et. al., 1998) preset a study of the use of MPI in for the Jacobi method of solving differential equations. Results of performance tests indicate that speedups of better than p=2 are possible with an optimal number of p nodes on a single Ethernet bus. Che et. al. (Che et al, 2004) propose an inter-nest cache reuse optimization method for Jacobi codes. This method is easy to apply, but effective in enhancing cache locality of the Jacobi codes while preserving their coarse grain parallelism.

This paper presents three algorithms for the parallelization of the Jacobi algorithm for the solution of linear equation systems. The main goal of these algorithms is to reduce the communication between the processors of the cluster, in order to increase the speed-up.

## 2  THE JACOBI ALGORITHM

The Jacobi algorithm is a well known numerical method used to solve linear algebraic systems of n equations with n unknowns. In this method, an initial approximate solution $X^0$ is selected and through an iterative procedure the algorithm tries to find the real solution **X**. If this solution exists, it is reached as the limit of a sequence of consecutive approximations $X^1, X^2, \ldots, X^m, \ldots$. This iterative procedure terminates when the algorithm reaches the actual solution, or when the predefined number of iterations has been exhausted.

In a more mathematical description let us consider an algebraic system described by the equations

$$\alpha_{11} x_1 + \alpha_{12} x_2 + \ldots + \alpha_{1j} x_j + \ldots + \alpha_{1n} x_n = \beta_1$$
$$\alpha_{21} x_1 + \alpha_{22} x_2 + \ldots + \alpha_{2j} x_j + \ldots + \alpha_{2n} x_n = \beta_2$$
$$\ldots \ldots \ldots \ldots \ldots \ldots \ldots$$
$$\alpha_{i1} x_1 + \alpha_{i2} x_2 + \ldots + \alpha_{ij} x_j + \ldots + \alpha_{in} x_n = \beta_i$$
$$\ldots \ldots \ldots \ldots \ldots \ldots \ldots$$
$$\alpha_{n1} x_1 + \alpha_{n2} x_2 + \ldots + \alpha_{nj} x_j + \ldots + \alpha_{nn} x_n = \beta_n$$

which can be combined to a single matrix equation **AX=B** where

$$A = \begin{pmatrix} \alpha_{11} & \alpha_{12} & \cdots & \alpha_{1j} & \cdots & \alpha_{1n} \\ \alpha_{21} & \alpha_{22} & \cdots & \alpha_{2j} & \cdots & \alpha_{2n} \\ \cdots & \cdots & \cdots & \cdots & \cdots & \cdots \\ \alpha_{i1} & \alpha_{i2} & \cdots & \alpha_{ij} & \cdots & \alpha_{in} \\ \cdots & \cdots & \cdots & \cdots & \cdots & \cdots \\ \alpha_{n1} & \alpha_{n2} & \cdots & \alpha_{nj} & \cdots & \alpha_{nn} \end{pmatrix}, \quad X = \begin{pmatrix} x_1 \\ x_2 \\ \cdots \\ x_i \\ \cdots \\ x_n \end{pmatrix}, \quad B = \begin{pmatrix} \beta_1 \\ \beta_2 \\ \cdots \\ \beta_i \\ \cdots \\ \beta_n \end{pmatrix}$$

are the matrix of coefficients, the solution of the system and the column-matrix with the constant terms respectively. The main idea behind the Jacobi method is to transform this system into an equivalent one of the form **X=TX+C**. To do this, we first write the matrix **A** as a sum of a diagonal matrix **D**, an upper triangular matrix **L** and a lower triangular matrix **U**, namely,

$$A = \begin{pmatrix} \alpha_{11} & 0 & \cdots & 0 \\ 0 & \alpha_{22} & \ddots & 0 \\ \vdots & \ddots & \ddots & 0 \\ 0 & \cdots & 0 & \alpha_{nn} \end{pmatrix} - \begin{pmatrix} 0 & 0 & \cdots & 0 \\ -\alpha_{21} & \vdots & & 0 \\ \vdots & \vdots & \vdots & \vdots \\ -\alpha_{n1} & \cdots & -\alpha_{n,n-1} & 0 \end{pmatrix} - \begin{pmatrix} 0 & -\alpha_{12} & \cdots & -\alpha_{1n} \\ \vdots & \ddots & & 0 \\ \vdots & & \vdots & -\alpha_{n-1,n} \\ 0 & \cdots & \cdots & 0 \end{pmatrix}$$

In this way, the equation **AX+B** can be written as **(D-L-U)X=B** or equivalently **DX=(L+U)X+B**. Therefore, if $\alpha_{ii} \neq 0$ (i=1,2,…,n) the inverse matrix **D$^{-1}$** exists and the solution vector **X** can be estimated as **X=D$^{-1}$(L+U)X+D$^{-1}$B**. From the last equation it is clear that the solution vector **X** can be estimated numerically by selecting an initial estimate **X$_0$** and by using the iterative equation **X$^{(k)}$=D$^{-1}$(L+U)X$^{(k-1)}$+D$^{-1}$B=TX$^{(k-1)}$+C** (k=1,2,..) where **T=D$^{-1}$(L+U)** and **C=D$^{-1}$B**.

Even though the sequence of the consecutive vectors produced in this way does not necessarily converge to some limit, it can be proven that this limit exists if $|\alpha_{ii}| > \Sigma_i |\alpha_{ij}|$. This limit is considered to be reached if the Euclidean distance between two consecutive vectors **X$^{(k)}$** and **X$^{(k+1)}$** falls below a predefined user supplied threshold. Since, in general, it is possible for the algorithm not to converge at all, the iterative procedure stops if the number of epochs specified by the user has been exhausted.

From the above description it is clear that the algorithm will fail if $\alpha_{ii}=0$ for any i. However, if the system has a unique solution, the equations can be reordered such that $\alpha_{ii} \neq 0$. In this case, the i$_{th}$ component of the vector **X$^{(k)}$** can be estimated by the equation

$$x_i^{(k)} = \frac{1}{\alpha_{ii}} \{ \sum_{j=1, j \neq i}^{n} (-\alpha_{ij} x_j^{(k-1)}) + \beta_i \} \quad (i = 1,2,...,n)$$

The pseudocode representing the Jacobi algorithm is presented below:

Input: (a) n×n matrix of coefficients such as $\alpha_{ii} \neq 0$, (b) n×1 vector of constant terms, (c) the initial approximation $X^t = X^0$, (d) the tolerance tol and the maximum iteration number L.

1) k=1
2) while k ≤ L do

   (a) for I=1,2,…,n, $x_i = -\sum_{j=1, j\neq i}^{n}(\alpha_{ij} X_j^t + \beta_i)/\alpha_{ii}$

   (b) if $\|X-X^t\|<$tol then return **X** and stop

   (c) k = k + 1; (d) $X^t = X$

3) Terminate the procedure if the iteration limit has been reached

The pseudocode presented above describes the serial implementation of the Jacobi algorithm running in a single CPU. However this algorithm can be easily parallelized since it contains matrix-vector multiplications that are highly parallelizable procedures. The description, the simulation and the evaluation of typical methods of parallelizing the Jacobi algorithm are presented in the next sections.

## 3 PARALLELIZING THE JACOBI ALGORITHM

To efficiently parallelize the Jacobi algorithm, the devised schemes should achieve data locality, minimize the number of communications, and maximize the overlapping between the communications and the computations. By assuming, for simplicity, that the number p of processes divides exactly the dimension n of the nxn matrix **A** and the vectors **X** and **B**, the distribution of the cells can be performed in a row-wise or a column-wise fashion. In the first approach, blocks of m=n/p rows of the **A** matrix are distributed to the system processes, while, vectors **X** and **B** are scattered in the same way. Regarding the second approach, it uses the same distribution scheme for the two vectors, but the matrix **A** is scattered not in rows, but in columns (namely, each process gets a set of m consecutive columns).

### 3.1 Row-wise data distribution

In the row-wise data distribution, only a part of the vector **X** is available on each process, and therefore, the matrix-vector multiplication cannot be carried out directly. To form the whole solution vector for each iteration of the algorithm, the MPI_Allgather function can be used to collect the partial vectors from all processes and concatenate them to form the whole vector; however this approach requires a lot of communications and it does not scale well for a large number of processors. A more appropriate parallelization scheme is based on the cyclic shift of the cells of the vector **X** in the way shown in Figure 1.

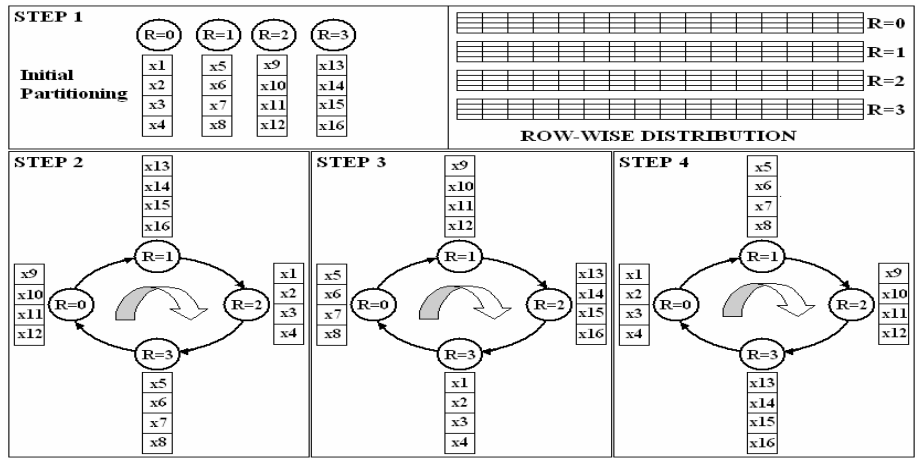

**FIGURE 1.** Cyclic shift of data in the row-wise distribution

From Figure 1 it is clear that in each step the partial vector is shifted upwards, and then, the partial matrix-vector multiplication is performed; in this way the multiplication **AX** is completed in p steps where p is the number of processes of the parallel application. Since in this approach each process needs to communicate only with the upper (for sending) and the lower (for receiving) neighbouring processes, the communication cost is less than the one associated with the MPI_Allgather - based approach. Furthermore, if non-blocking requests for sending and receiving are used, the overlapping between computations and communications can be achieved. Actually, since these requests return immediately, each process can perform the partial computations with the available data during the data transfer operations, and check periodically the non-blocking communication status by using the MPI_Test function (in cases where the computations are finished before the termination of the data transfers, the process should wait for them, by calling the MPI_Wait function).

In a mathematical description, the process with rank R ($0 \leq R \leq p-1$) estimates the partial vector $\mathbf{X}_p = \{X_{Rm}, X_{Rm+1}, X_{Rm+2}, \ldots, X_i, \ldots, X_{(R+1)m-1}\}$. In each one of the p steps, only a part of the value of each component is estimated; more specifically, the partial value of the $X_i$ component [$Rm \leq i \leq (Rm+m-1)$] estimated during the $s_{th}$ step [$0 \leq s \leq (p-1)$] is given by the equation

$$X_{i(s)}^{(k+1)} = \sum_{j=s*m}^{(s+1)m-1} \alpha_{ij} x_j^{(k)}$$

Regarding the final value of the $X_i$ component [$Rm \leq i \leq (R+1)m-1$] it can be estimated by adding all these partial values and therefore it is given by the equation

$$X_i^{(k+1)} = \sum_{s=0}^{p-1} X_{i(s)}^{(k+1)} = \sum_{s=0}^{p-1} \{ \sum_{j=s*m}^{(s+1)m-1} \alpha_{ij} x_j^{(k)} \}$$

Finally, to check if the stopping condition

$$\left\| X^{(k+1)} - X^{(k)} \right\| = \sqrt{\sum_i (X_i^{(k+1)} - X_i^{(k)})^2} < tol$$

holds, a global data gathering operation is needed that can be easily performed by calling the MPI_Gather function and choosing one of the processes (for example the process R=0) as the root process for this operation.

### 3.2 Column wise data distribution

In the column wise data distribution, each process holds a block of m columns and a partial **X** vector, and therefore, the partial product **AX** can be computed independently and concurrently with the other processes. This fact can be easily understood; the block of m consecutive columns forms a two dimensional sub-matrix with dimensions nxm, whose product with the partial vector - with dimensions mx1 - leads to a vector with dimensions nx1. This means that each process estimates a partial value of all the cells of the **X** vector - since it uses in this matrix-vector multiplication only its own columns - and therefore to get the final value, all these partial values estimated by all processes have to be added together and subtracted by the value of the corresponding cell of the **B** vector. An obvious way to do this, is to call the MPI_Allreduce function with the MPI_SUM opcode argument to perform a global reduction, but the most preferable way is again the cyclic shift, which in this case, is performed as follows:

The coefficient matrix is shifted leftwards while the vector of unknowns is shifted upwards and then, the partial product **AX** is estimated with the result to be subtracted by the vector **B**. This operation is performed in p steps. Then, the whole product **AX** is be estimated and subtracted by the vector **B**, thus providing the solution vector **X**. The operation of the cyclic shift for the column wise distribution is shown in Figure 2 (the cyclic shift of the solution vector is performed in the same way as in the row wise operation and it is shown in Figure 1).

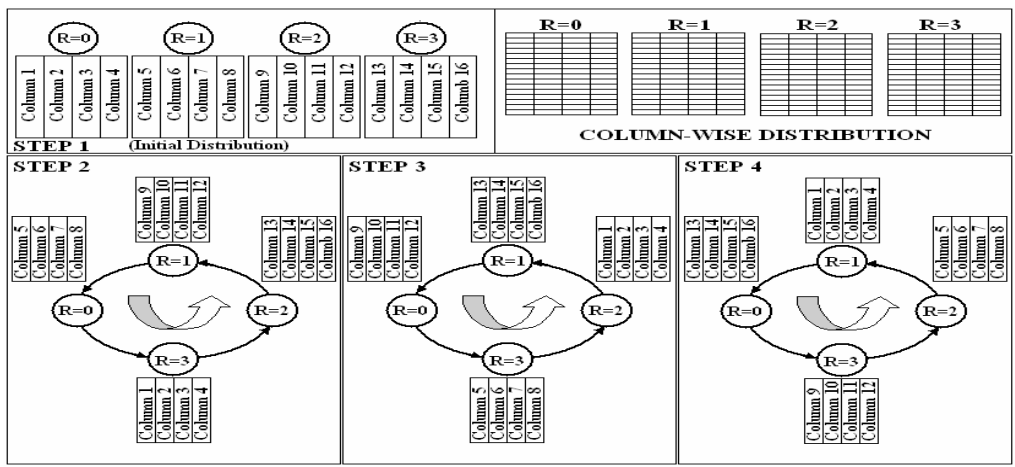

**FIGURE 2.** Cyclic shift of data in the column-wise distribution

To present again a mathematical description, we can easily note that the process R [$0 \leq R \leq (p-1)$] holds during the execution of the $s_{th}$ step [$0 \leq s \leq (p-1)$] the columns with indices [$(R+s)m$ mod $n$],[$(R+s)m$ mod $n$]+1, … ,[$(R+s)m$ mod $n$]+m-1 where mod is the modulo operator; therefore, the partial value of the component $X_i$ ($0 \leq i \leq n$) estimated during the $s_{th}$ step is given by the equation

$$X_{i(s)}^{(k+1)} = B_i - \sum_{j=[(R+s)m \bmod n]}^{[(R+s)m \bmod n]+m-1} \alpha_{ij} x_j^{(k)}$$

(where $B_i$ is the value of the corresponding cell of the B vector), while, the total value of the component $X_i$ ($0 \leq i \leq n$) is estimated as

$$X_i^{(k+1)} = B_i - \sum_{s=0}^{p-1} X_{i(s)}^{(k+1)} = B_i - \sum_{s=0}^{p-1} \{ \sum_{j=[(R+s)m \bmod n]}^{[(R+s)m \bmod n]+m-1} \alpha_{ij} x_j^{(k)} \}$$

From the above description it is clear that the $\mathbf{X}^{(k+1)}$ vector is estimated in each process and therefore the validity of the stopping condition can be checked by one of the processes (for example by the process R=0). We note that the column-wise method requires much more communication than the row-wise one, since, besides the upwards shifting of vector **X**, the coefficient matrix **A** is also shifted leftwards. However, in this approach, the global reduction operation is not needed, since a partial value of all cells is estimated at each step. Therefore there is not a direct way to compare the efficiency of the two methods.

**3.3 One-sided implementation of the Jacobi algorithm**

An alternative, quite different parallelization scheme for the Jacobi algorithm can be implemented by using the one-sided communication functions provided by the MPI library that allow the usage of remote memory access operations. The main idea of this new approach is to use an additional target process to store the vectors $\mathbf{X}^{(k)}$ and $\mathbf{X}^{(k+1)}$, and estimate the distance of their current instances, thus checking for the terminating condition. Therefore, the parallel application consists of p+1 processes with the first p processes with ranks {0,1,2,….,p-2} used as before, while the last (additional) process with rank R=p-1 is the target process of the application. The vectors $\mathbf{X}^{(k)}$ and $\mathbf{X}^{(k+1)}$ are stored in the appropriate memory windows created in the memory of the target process, and their contents are read and written by the other processes utilizing the MPI_Get and MPI_Put functions respectively. After the estimation of the whole $\mathbf{X}^{(k+1)}$ vector, the distance D=||$\mathbf{X}^{(k+1)}$-$\mathbf{X}^{(k)}$|| is estimated by the target process, and if it is less than the user supplied threshold, the algorithm terminates. Otherwise, the vector $\mathbf{X}^{(k+1)}$ is copied into the memory window that keeps the old vector $\mathbf{X}^{(k)}$; this copy operation is performed locally in the memory of the target process without the requirement of communication with the remaining system processes. The process communication in the one-sided Jacobi algorithm implementation is shown in Figure 3.

The synchronization of the processes in this approach can be achieved by calling the functions MPI_Win_post, MPI_Win_wait, MPI_Win_start, and MPI_Win_complete. Therefore, the target process is considered to be an active process, while, alternative implementations can be designed based to the passive target process approach and to the use of functions MPI_Win_lock and MPI_Win_unlock. To use the synchronization functions, two process groups has to be created; the origin group containing the processes with ranks {0,1,2,…,p-2} and the target group that contains the last target process. The process synchronization scheme described above is shown in Figure 4.

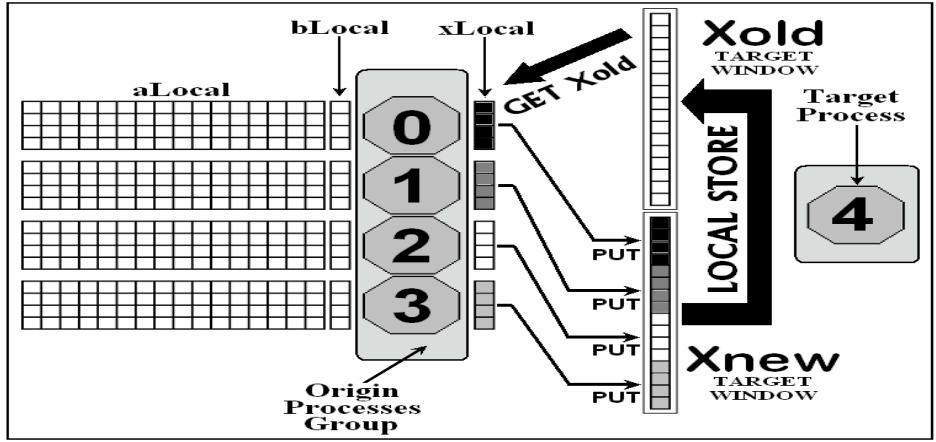

**FIGURE 3.** Data communication of the one-sided algorithm

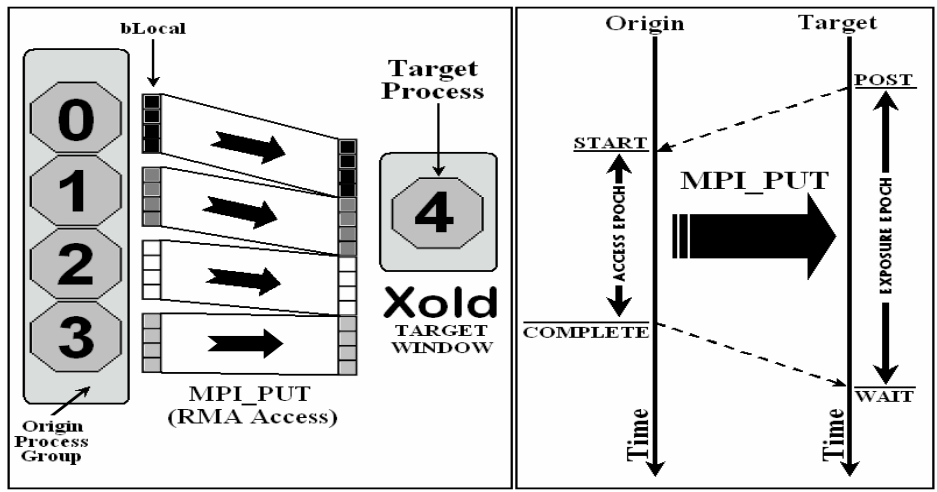

**FIGURE 4.** Process synchronization in one-sided implementation of the Jacobi algorithm

Based on the above description, the one-sided based Jacobi algorithm implementation is composed by the following steps:

(1) Root process (R=0) initializes the matrix **A** and the vectors **X** and **B** and distributes them to all system processes except the target one by using row-wise distribution.

(2) Target process (R=p-1) calls the MPI_Win_create function to create the memory windows used for the storage of the vectors $X^{(k)}$ and $X^{(k+1)}$. (Note: since this is a collective function it will be called by all processes; however all processes different from the target, will call this function with the MPI_BOTTOM as the window argument, to prevent the creation of the window in their memories).

(3) The origin and the target process groups are created; then, each process stores its partial vector to the appropriate position of the $X^{(k)}$ memory window in the memory of the target process by using the MPI_Put function.

(4) For each cycle k ($0 \leq k \leq L-1$)
- Each non target process calls the MPI_Get function to get the $X^{(k)}$ vector from the appropriate target window
- Each non target process estimates its partial $X^{(k+1)}$ vector and stores it to the appropriate cells of the associated memory window in the target process memory.
- The target process estimates the distance D between these two vectors and broadcasts its value to the processes of the origin group. If this distance is less than the user supplied tolerance, the algorithm terminates.

The process synchronization in the above communications is guaranteed by using the functions MPI_Win_post and MPI_Win_wait in the target process side and the functions MPI_Win_start and MPI_Win_complete in the origin process side.

## 4 EXPERIMENTAL RESULTS

The experimental results presented below, came from simulations of the parallel algorithms described above; these simulations were run on the Electra grid system of the Department of Applied Informatics in the University of Macedonia. The grid was created in 2003 by the Parallel and Distributed Processing Laboratory and it is composed of the Electra server and sixty five computing nodes connected to the server via a two level tree structure of 100 MBps Ethernet switches. The operating system software is based on RedHat Linux, NPACI Rocks and Ganglia toolkit, and the installed applications include C and Fortran compilers, the MPICH implementation of the MPI library and a set of useful parallel libraries such as Blas, Lapack and Scalapack.

To measure the speedup of the parallel application, the Jacobi algorithm was implemented for solving linear systems with dimensions n=8,16,32,64,128,256,512. Since the objective of the research was only to measure the speedup and not to actually solve the system, the cells of the matrix **A** and the vectors **B** and **X** were initialised to small random values in the interval [0,1], while the tolerance parameter was set to the value of $10^{-8}$ to guarantee that the number of iterations L will be reached (in these simulations the value of L=10000 was used). The presented results are the execution times of the parallel application with p=2,4,8,16,32 processes for the row-wise and the column-wise approaches, and for the blocking and non-blocking communication modes. The one-sided implementation of the Jacobi algorithm could not be tested; the set of MPI functions used in its implementation are part of the MPI2 library, a library that is not supported by the Electra grid, that runs the MPI1 message passing library. The simulation of this implementation and the measure of its performance is a part of the future work in this area.

**TABLE 1.** Simulation results for the serial case (p=1)

| Dimension N | Execution time t (in seconds) |
|---|---|
| 008 | 0001.497 |
| 016 | 0004.478 |
| 032 | 0017.430 |
| 064 | 0069.501 |
| 128 | 0279.694 |
| 256 | 1097.305 |
| 512 | 4386.621 |

**TABLE 2.** Simulation results for the blocking row-wise distribution

| DIM = 8 | | DIM = 16 | | DIM = 32 | | DIM = 64 | |
|---|---|---|---|---|---|---|---|
| P=02 | T=006.015 | P=02 | T=006.740 | P=02 | T=007.940 | P=02 | T=015.721 |
| P=04 | T=012.762 | P=04 | T=013.098 | P=04 | T=013.646 | P=04 | T=017.023 |
| P=08 | T=024.981 | P=08 | T=023.645 | P=08 | T=024.088 | P=08 | T=026.505 |
| P=16 |  | P=16 | T=046.814 | P=16 | T=046.503 | P=16 | T=048.005 |
| P=32 |  | P=32 |  | P=32 | T=132.126 | P=32 | T=138.225 |

| DIM = 128 | | DIM = 256 | | DIM = 512 | |
|---|---|---|---|---|---|
| P=02 | T=015.393 | P=02 | T=164.468 | P=02 | T=642.823 |
| P=04 | T=016.846 | P=04 | T=092.057 | P=04 | T=331.323 |
| P=08 | T=028.653 | P=08 | T=065.657 | P=08 | T=185.699 |
| P=16 | T=049.796 | P=16 | T=069.060 | P=16 | T=134.919 |
| P=32 | T=130.521 | P=32 | T=148.243 | P=32 | T=190.299 |

Typical simulation results for the row-wise distribution are presented in Figure 5. In this figure the curve with circular markers corresponds to the results for blocking mode, while the curve with square markers corresponds to the results for the non-blocking mode. As can be seen, for linear systems of small size (8 or 16) the communication overhead is large compared to the computation time, which leads to speed-ups less than one. However, for large systems (dimensions grater than 32) the speed-up is quite good, even more than 30 for a system of dimension 512 divided into 16 processes. The results for the column-wise speed-up were not plotted, since, as can be seen in Table 2, the execution times are worse than the execution times on the serial system, which leads to speed-up less than one. It is obvious that the column-wise distribution suffers from a large communication overhead, which makes it slower then the serial system at least for vector dimensions up to 128.

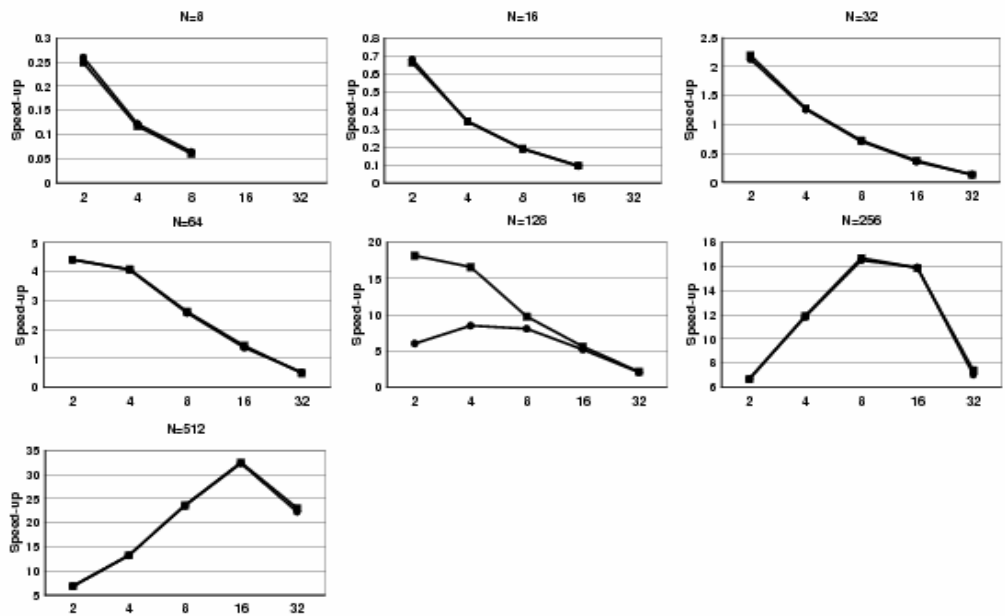

**FIGURE 5.** Simulation results for the row-wise distribution

## 5  CONCLUSIONS

The inherent division of the Jacobi algorithms into smaller tasks offers a good candidate for parallelizing these algorithms on an MPI cluster. However, due to high data dependencies, the parallelization algorithms should be designed in such a way as to minimize the communication overhead, which can severely affect the performance. As shown in this paper the column-wise distribution of the data sets among the processors is not very efficient, since it does not eliminate most of the communication overhead. On the other hand, the row-wise distribution gives very good results, reaching a speed-up of up to 32. What remains to be done in future work is the one-sided implementation of the Jacobi algorithm and its comparison to the row-wise distribution.

**REFERENCES**

1   Acacio M et. al (1998). The Performance of MPI Parallel Jacobi Implementation on Workstation Clusters, *IX Jornadas de Paralelismo*, p.261.
2   Che Y, Wang Z, Li X, Yang L.T (2004.) Locality Optimizations for Jacobi Iteration on Distributed Parallel Systems, *Lecture Notes in Computer Science*, Volume 3358, p. 91.

3  Coope I.D, Macklem M.S (2005). Parallel Jacobi methods for derivative-free optimization on parallel or distributed processors, *Australian and New Zealand Industrial and Applied Mathematics Journal*, Vol.46, p. 719.
4  Gimenez D, van de Geijn R, Hernandez V, Vidal A.M (1995). Exploiting the Symmetry on the Jacobi Method on a Mesh of Processors, in *Proceedings of the 4th Euromicro Workshop on Parallel and Distributed Processing (PDP '96)*, p. 377.
5  Londre T, Rhee N.H (2005). Numerical Stability of the Parallel Jacobi Method, *SIAM Journal on Matrix Analysis and Applications*, Vol.26, No. 4, p. 985.
6  Luke F.T., Park H (1989). A Proof of Convergence for Two Parallel Jacobi SVD Algorithms, *IEEE Transactions on Computers*, Vol.38, No. 6, p. 806.
7  Royo D, Gonzalez A, Valero-Garcia M (1998). Jacobi Orderings for Multi-Port Hypercubes, in *Proceedings of the 12th. Intern. Parallel Processing Symposium*, p. 88.
8  Zhou B.B, Brent R.P (1995). On parallel implementation of the one-sided Jacobi algorithm for singular value decompositions, in *Proceedings of the 3rd Euromicro Workshop on Parallel and Distributed Processing*, p. 401.